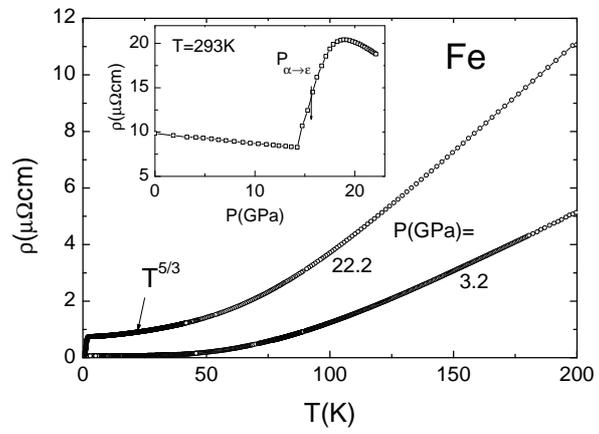

Jaccard_fig2

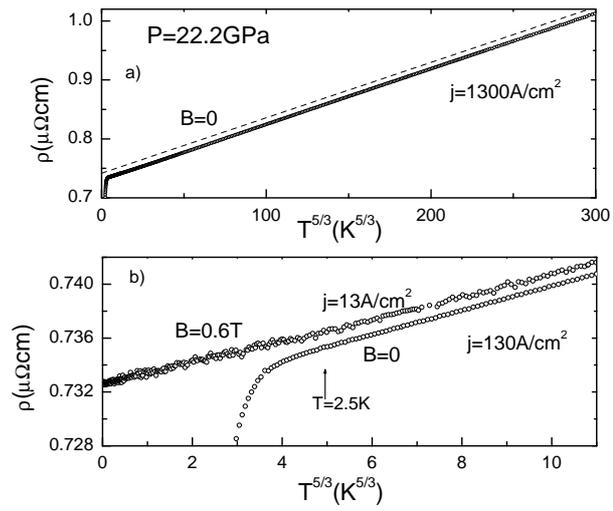

Jaccard_fig3

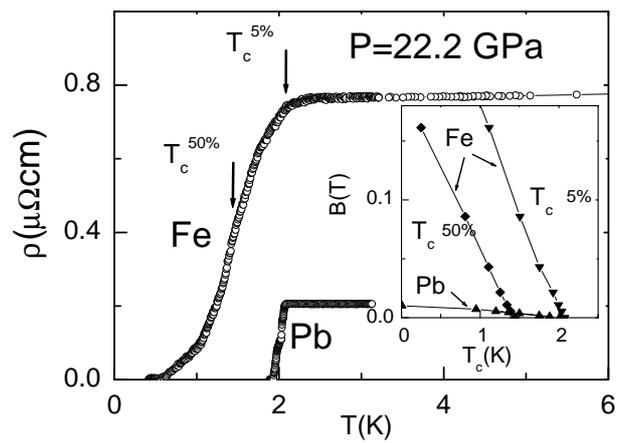

Jaccard_fig4

# Superconductivity of ε-Fe: complete resistive transition

D. Jaccard [a,*], A. T. Holmes [a], G. Behr [b], Y. Inada [c], Y. Onuki [c]

[a] *DPMC, University of Geneva, 24 Quai Ernest-Ansermet, 1211 Genève 4, Switzerland*
[b] *Institute for Solid State and Materials Research Dresden, BP 270016, D-01171 Dresden, Germany*
[c] *Department of Physics, Graduate School of Science, Osaka University, Osaka 560-0043, Japan*

**Abstract**

Last year, iron was reported to become superconducting at temperatures below 2K and pressures between 15 and 30 GPa [1]. The evidence presented was a weak resistivity drop, suppressed by a magnetic field above 0.2 T, and a small Meissner signal. However, a compelling demonstration, such as the occurrence of zero resistance, was lacking. Here we report the measurement of a complete resistive transition at 22.2 GPa with an onset slightly above 2 K in two very pure samples of iron, of different origins. The superconductivity appears unusually sensitive to disorder, developing only when the electronic mean free path is above a threshold value, while the normal state resistivity is characteristic of a nearly ferromagnetic metal.

* Corresponding author.
  *E-mail address:* Didier.Jaccard@physics.unige.ch
  *Postal address*: D. Jaccard,
  DPMC, University of Geneva, 24 Quai Ernest-Ansermet, 1211 Genève 4, Switzerland
  Phone: 41 22 702 63 66,   Fax: 41 22 702 68 69

Two iron samples and one lead strip were mounted in series in a pressure cell for sensitive electrical resistivity measurements (fig. 1). The high pressure cell [2] is made from a pyrophylitte gasket containing a steatite (soapstone) pressure transmitting medium. Both the gasket and the transmitting medium are squeezed between the flats of two opposed anvils made of sintered diamond. The load on the anvils, and hence the pressure in the cell, is changed at room temperature and stays constant when the temperature is lowered. The pressure is deduced from the temperature of the superconducting transition of the lead strip. The width of this transition indicates the pressure gradient $\Delta P$ – at 22 GPa, $\Delta P$ = 0.85 GPa (4% of the applied pressure). The setup shown in Fig. 1 allows us to perform accurate resistivity measurements at high pressure using the four-point method. The absolute resistivity can be determined to within 10%, the main error being on the sample cross-section.

The two iron samples shown in fig.1 and labelled Fe # 1 and Fe # 2 in the following have been cut from two different very pure batches. The purification of sample #1 is described in ref.[1], while sample #2 came from a rod composed of large single crystalline grains grown in Dresden 30 years ago [3]! The main impurities in #2 are Cu (~14 ppm), Mn (1.6 ppm), O (18 ppm) and N (<5 ppm). All other metallic impurities are below or far below 1 ppm (e.g. Ni<0.84 ppm, Co~0.18 ppm, Cr~0.13 ppm). The residual resistivity ratio, RRR = $\rho$(293 K) / $\rho(T \rightarrow 0$ K), a measure of sample quality, also depends on defect concentration. The RRR fell from an initial value of 200-250 in both batches, to RRR ≈ 160, when the sample was cut to a cross-section compatible with our pressure cell (~30 μm × 15 μm), despite our efforts to minimise the cold work during this process.

At ~14 GPa, iron undergoes a first order transition from the ferromagnetic bcc α-phase found at ambient pressure, to the hcp ε-phase, believed to be nearly magnetic. It is in this ε-phase that superconductivity was discovered by the Osaka team [1]. The onset of the bcc to hcp transition can be seen in the inset of fig. 2 as a sharp kink in $\rho(P)$ at 293 K. The increase of $\rho$ reflects the change in the electron-phonon scattering at $P_{\alpha \rightarrow \varepsilon}$. The transition is sluggish, observed over a much broader pressure range than the pressure gradient $\Delta P$, and pressures higher than 20 GPa are needed for a complete transformation.

At 3.2 GPa iron is in its low pressure phase. Fig. 2 shows that the resistivity, $\rho(T)$, is characteristic of ferromagnetic iron at this pressure. Above ~100 K, slightly superlinear resistivity is found, due to electron-magnon scattering superimposed on the standard linear electron-phonon term. Below ~ 30 K, $\rho(T)$ approximately follows the Fermi-liquid law $\rho(T) = \rho_0 + AT^2$, where $\rho_0 = 0.06$ $\mu\Omega$cm is the residual resistivity and $A = 2\times10^{-5}$ $\mu\Omega$cmK$^{-2}$. The coefficient $A = 2\times10^{-5}$ $\mu\Omega$cmK$^{-2}$ is very close to the value observed at $P = 0$ [4]. It is mainly due to s-d interband electron scattering and reflects the moderately large 3d-electron density of states at the Fermi level. A good fit up to 50 K is found by introducing an additional textbook $T^5$ low-temperature electron-phonon contribution.

At 22.2 GPa, iron is in its hcp $\varepsilon$-phase. Below room temperature the resistivity $\rho(T)$ is linear with no additional terms, since the long range ferromagnetic order has vanished. Three remarkable features emerge at low temperature: Firstly, the residual resistivity jumps by more than an order of magnitude at the structural transition and for sample # 1 at 22.2 GPa, $\rho_0 = 0.76$ $\mu\Omega$cm (fig. 2). It seems that defects are introduced at this point. Secondly, below 30 K $\rho(T)$ varies as $bT^{5/3}$ (shown in fig. 3), not the quadratic dependence expected for a conventional Fermi liquid. This has never been reported for a pure metallic element before. The large $b$ coefficient points to a strong enhancement of electronic correlations. Finally, below 2.5 K a complete superconducting transition is observed.

Figure 4 shows the superconducting transition of $\varepsilon$-Fe at 22.2 GPa. The transition is broad, with an onset just above 2 K, and completion at about 0.5 K. In contrast, the lead manometer exhibits a much narrower transition, which reflects the pressure gradient in the cell, with $\Delta P = 0.85$ GPa. 22.2 GPa corresponds to the maximum $T_c$ found in ref. [1], implying that the width of the Fe transition is not due to $\Delta P$. The upper limit on $\rho$ for the completed transition, below 0.5 K, is 3 n$\Omega$cm, i.e. the resistance has dropped by more than a factor of 230 with a current density of $j = 0.25$ A/cm$^2$. With a larger $j$, the transition becomes incomplete, but a 50% drop in $\rho$ still occurs for $j$ as high as 1300 A/cm$^2$. Surprisingly, the $T_c$ onset does not depend so strongly on the current. In fig. 3b, where $j = 130$ A/cm$^2$, $T_c^{onset}$ is close to 2.5 K.

Application of a magnetic field $B$ provides further information. A field of only 0.02 T prevents the completion of the superconducting transition, indicating that part of the sample exhibits a low critical field $B_c$. Other criteria lead to much higher estimates of $B_c$. $T_c^{5\%}$, defined by a 5% resistivity drop, leads to $B_c(T_c = 0) \approx 0.3$ T; even higher critical fields can be found if $T_c^{onset}$ is considered. $T_c^{5\%}$ and $T_c^{50\%}$ are shown in the inset of fig. 4. A comparison with the lead transition shows that $B_c$ of $\varepsilon$-Fe is at least 30 times larger than that of Pb (for a similar $T_c$ at $B = 0$). The initial slope is twice that of Nb. Such a high critical field suggests that $\varepsilon$-Fe is a type II superconductor. The corresponding superconducting coherence length $\xi$ is about 30 nm, implying a reduced Fermi velocity (assuming the clean limit), and hence that the effective electron mass is enhanced by a factor of about six with respect to $\alpha$-Fe.

The results for sample #2 are very similar. A complete resistive transition is also obtained, but only for a slightly lower current density $j$ (< 0.15 A/cm$^2$). $T_c^{5\%} = 1.86$ K, a value marginally smaller than for sample #1, and $B_c$ is reduced by about 30%, while the residual resistivity is quite similar ($\rho_0 = 0.73$ $\mu\Omega$cm).

If the superconductivity is completely suppressed with a field of 0.6 T, the resistivity broadly keeps its $T^{5/3}$ behaviour below 2.5 K (fig. 3). Above 2.5 K the resistivity is simply shifted by a positive magnetoresistance of 0.1% with respect to the zero field value. The effective exponent $n(T)$ can be found using a local fit to $(\rho - \rho_0) \propto T^{n(T)}$. When $4 < T < 30$ K, $n = 1.67 \pm 0.05$ is obtained for both samples. Above 30 K, $n$ increases slightly due to the $T^5$ phonon term and, at still higher temperature, $n$ decreases to 1 as expected. Below 4 K, there is a clear trend for $n$ to increase as $T \to 0$ K. This trend is more pronounced in sample #1 and, at about 2 K, the Fermi-liquid value $n = 2$ is nearly reached, with a temperature coefficient $A$ in the range of $6\times10^{-4}$ $\mu\Omega$cmK$^{-2}$ – 30 times larger than for $\alpha$-Fe.

A $T^{5/3}$ power law in resistivity is predicted by the nearly ferromagnetic Fermi-liquid model [5, 6, 7]. This model is an extension of the standard Fermi-liquid theory of the metallic state when weak ferromagnetic correlations are considered among itinerant electrons. The model predicts a $T^{5/3}$ power law at low temperature in a 3D system close to a magnetic instability with moderate impurity scattering. Below a certain temperature $T^*$, the resistivity should recover a $T^2$ dependence down to $T \to 0$. We found $T^*$ to be surprisingly low, which could indicate that the system is close to a quantum critical point. However, there is no obvious reason to expect such an instability at 22.2 GPa. An explanation for this behaviour might lie in the sluggish nature of the $\alpha \to \varepsilon$ structural transition (depending notably on the pressure medium, as revealed by Mössbauer effect experiments [8]). This is consistent with our room temperature data (inset fig. 1). A small amount of $\alpha$-Fe likely persists at our measuring pressure; unstable ferromagnetic clusters could strongly influence the electronic transport causing our samples to behave like a nearly ferromagnetic metal. Mössbauer measurements show that the magnetic susceptibility is strongly enhanced to nearly $10^{-2}$ emu/mol [9], corresponding qualitatively to our finding a large temperature coefficient of resistivity [10], associated with the enhanced effective mass. On the other hand, the persistence of $T^{5/3}$ resistivity down to very low temperature (i.e. low $T^*$) points to a breakdown of Fermi-liquid theory, a phenomenon that has recently commanded much attention. This breakdown is found in other systems, such as heavy fermion alloys, where disorder has been proposed to induce deviations from Fermi-liquid behaviour over a wide pressure range [11]. By analogy, it may be that magnetic clusters in $\varepsilon$-Fe are at the origin of the intriguing persistence of the $T^{5/3}$ resistivity.

Superconductivity in $\varepsilon$-Fe appears to be strongly sensitive to disorder [1,12]. No trace of superconductivity was found in a previous sample with similarly high purity of 99.998% (Goodfellow). This was simply rolled and then cut with a razor blade to fit into the pressure cell. As a result, the residual resistivity was relatively large, with values of 0.6 $\mu\Omega$cm and 2 $\mu\Omega$cm found in the $\alpha$ and $\varepsilon$-phases respectively. Moreover, we found that the cold work experienced by a high purity sample largely determines its residual resistivity. Superconductivity was found to develop when $\rho_0 < 2$ $\mu\Omega$cm, i.e. the electronic mean free path is larger than a threshold value. If samples with even lower residual resistivities can be produced, our results and those from ref. [1] would lead one to expect narrower superconducting transitions and an even higher $T_c$. The mean free path $\ell$ can be estimated by the standard treatment of resistivity (through the Boltzmann equation) using recent band structure calculations [13]. One obtains $\ell$ = 12 nm [14], not very different from the superconducting coherence length $\xi$. It may be that the clean limit $\ell > \xi$ is required for the appearance of superconductivity.

Superconductivity in $\varepsilon$-Fe has now been firmly established. The next challenging question is the nature electron pairing mechanism. The first principle calculations in ref. [13] show that superconductivity mediated by ferromagnetic fluctuations is a strong possibility for $\varepsilon$-Fe. The low temperature $T^{5/3}$ resistivity observed lends support to this scenario. A nearly antiferromagnetic ground state, often envisaged for $\varepsilon$-Fe [15, 16], would imply a $T^{3/2}$ resistivity law, clearly at odds with our data. Understanding $\varepsilon$-Fe is not only essential for theories of superconductivity but also has geophysical interest: The Earth's solid inner core is composed primarily of iron, and the magnetic properties of $\varepsilon$-Fe may play an important role in the planetary magnetic field [15, 17]. Finally, high sensitivity to non-magnetic disorder is a characteristic feature of spin-triplet superconductivity. Iron may to be the first pure element found to exhibit such unconventional superconductivity, and for that reason alone deserves further study.

**Acknowledgments**
We thank J. Flouquet for stimulating discussions.

**Figure captions:**

Figure 1. Sample chamber before pressurisation. The two iron samples are placed on disk of soft pressure medium (steatite, $\phi$ = 1mm), which is contained by an annular gasket (pyrophyllite). Thin Au wires ($\phi$ = 25 µm) placed on the samples and pressure gauge (Pb foil) provide connections across the gasket. A second steatite disk is placed on top of the cell which is then pressurised between two Bridgman anvils made of sintered diamond.

Figure 2. Temperature dependence of the resistivity $\rho$ of iron in the α and ε-phase. The inset shows the pressure dependence of $\rho$ at room temperature.

Figure 3. a, The resistivity of ε-Fe (sample # 2) on a $T^{5/3}$ scale below 30 K. Dashed line represents a $T^{5/3}$ behaviour. b, In a magnetic field $B$ = 0.6 T, the $T^{5/3}$ dependence persists below 2.5 K while superconductivity develops at $B$ = 0.

Figure 4. The superconducting transition of ε-Fe at 22.2 GPa (sample #1). Below 0.5 K, $\rho$ is less than 3 nΩcm. For comparison, the transition of the lead manometer is shown. The inset shows the critical field of ε-Fe for two $T_c$ criteria indicated by arrows in the main figure. For Pb, we used $T_c^{50\%}$. The results for Fe #2 are very similar with slightly reduced $T_c$ and $B_c$ values.